\documentclass[11pt]{article}
\usepackage{moriond,epsfig}

\def\Journal#1#2#3#4{{#1} {\bf #2}, #3 (#4)}

\def\NPB{{\em Nucl. Phys.} B}
\def\PLB{{\em Phys. Lett.}  B}
\def\PRL{\em Phys. Rev. Lett.}
\def\PRD{{\em Phys. Rev.} D}

\def\be{\begin{equation}}
\def\ee{\end{equation}}
\def\bea{\begin{eqnarray}}
\def\eea{\end{eqnarray}}

\begin{document}
\vspace*{4cm}
\title{MODELS AND SIGNATURES OF EXTRA DIMENSIONS AT THE LHC}

\author{ M. BESANCON }

\address{CEA-Saclay/Irfu/SPP, Bat.~141,\\
91191 Gif sur Yvette, France}

\maketitle\abstracts{
Models for extra dimensions and some of the most promising 
ensuing signals for experimental discovery at the LHC are 
briefly reviewed. The emphasis will be put on the production 
of Kaluza Klein states from both flat and warped extra-dimensions 
models.}

\section{Introduction}

One of the first motivation for the recent renewed interest for 
extra dimensions (since the early years of the 20th century, back in the 
G.~ Nordstr\"om~\cite{nord}, T.~Kaluza~\cite{kalu}, O.~Klein~\cite{klei}
A.~Einstein and P.~Bergmann~\cite{ein}
time where they were first discussed from the physics point of view~\footnote{Not
to mention the older mathematical point of view back in the 19th century with B.~Riemann
and G.~Cantor including among other 
the very  basic notion of what a dimension actually is and means which
turns out to be a non trivial question (G.~Cantor discovered in 1877 that the points of a square i.e.
2~dimensional, can be put into one to one correspondance with the points of a line segment 
i.e. one dimensional, {\it ``thus rendering the simple idea of dimension problematic''} - D. Johnson) 
also interesting to be discussed. For this latter notion, we refer the 
reader to the following (non exhaustive list) textbooks: Dimension Theory by  W.~Hurewicz and 
H.~Wallman, Princeton 1941 and Modern Dimension Theory by J.~Nagata, 
North-Holland, 1965.} and since some further developments from the 50's 
to the 70's) comes from the possibility they have to address the hierarchy 
problem of the standard model of particle physics. In the course of their 
development, it quickly appears that they can also address some other issues 
such as symmetry breaking, provide some understanding for masses and mixing of 
the Standard Model fermions, allow for TeV scale unification without supersymmetry 
and provide dark matter candidates. There are also strong motivations for extra 
dimensions from more fundamental underlying candidate theories such as for example 
string theories that will not be discussed in the following since it would bring 
this short review too far out of scope. 

There are many possible approaches for 
extra dimensions models connected to basic questions such as: how many extra 
dimensions can there be ? Which geometry can they have ? How large can they be, 
with which consequences for the phenomenology ? Which fields are sitting where ?

There are two kinds of possible geometries for
extra dimensions.
One is called factorizable or flat where one
can have in principle any number of extra dimensions i.e.
3~space dimensions + 1~time dimension + $D-4$ extra-dimensions, 
with a metric written in the
so called usual way $ds^2 = g_{\mu \nu} dx^{\mu} dx^{\nu}$ ($\mu, \nu = 0,1,2,3 ...D$).
The other one is called non factorizable or warped and is characterized by the presence 
of a warp factor $a(y)$, depending on usually only one extra-dimension~$y$, put in front 
of the 4~dimensional metric (often identified with the Minkowki metric $\eta_{\mu \nu}$)
$ds^2 = a(y) (\eta_{\mu \nu} dx^{\mu} dx^{\nu}) + dy^2$ (where here $\mu, \nu = 0,1,2,3$).

Extra dimensions have not yet been seen experimentally
so if they exist they must be small (for flat geometries) 
i.e. compact. Compactifying extra dimensions leads to
some periodicity conditions on fields 
so that one can Fourier expand them: 
\begin{equation}
\phi(x_{\mu},y) = 
\sum_{k=-\infty}^{+\infty} 
{\phi^{(k)}(x_{\mu})}
e^{{{iky}\over{R}}}
\label{eq:kakl}
\end{equation}
and separate the 4 dimensional component
which gives rise to the so called Kaluza-Klein (KK)
modes or states or excitations ${\phi^{(k)}(x_{\mu})}$. 
The number of KK states is infinite. They are massive
and the mass of the $k^{th}$ mode is given by the inverse 
compactification size/radius $R$:
\begin{equation}
m^2_k = m^2_o + {{k^2} \over {R^2}}.
\label{eq:kkm}
\end{equation}

Answering the question which field is sitting where leads to an immense variety of 
possible approaches for extra dimensions. One of the first and simplest approach is 
the so called ADD model~\cite{add} where only gravity propagates in the full D~dimensional 
spacetime with flat geometry and with n compactified extra space dimensions which will be 
called the 
bulk~\footnote{See also~\cite{koko} for non-supersymmetric string models that can realize extra-dimension
scenario and break to only the Standard Model without extra massless matter.}. 
The compactified extra dimensions can be quite large as we will see later 
on. The Standard Model fields are confined on a 4~dimensional sub-spacetime which will be 
called brane.
The so called  $TeV^{-1}$ models are models where one can have one or more small compactified
extra dimensions with flat geometries with sizes of the order of
the inverse of the TeV scale i.e. $O(10^{-19}) \, m$, where the Standard Model gauge
bosons can propagate ~\cite{tev}. This whole setup can be embedded in a larger
space where gravity propagates. The fermions of the Standard Model are still confined on a 
4~dimensional sub-spacetime. 
Universal extra-dimension (UED) models~\cite{ued} are models with flat geometry
where Standard Model gauge bosons as well
as fermions can propagate in the bulk with one (or more recently two) extra dimension. 
An important set-up is the so called RS setup~\cite{rs} which is a setup where gravity 
only propagates in a 5~dimensional warped bulk with one compactified extra dimensions 
and with two 4~dimensional branes. The Standard Model fields are confined 
on one brane i.e. the infrared or TeV brane, is at the TeV scale while the other brane, the
Planck brane, is at the Planck scale. One can further extend the previous minimal RS
setup by putting a scalar field in the warped bulk 
which will allow to stabilize the interbrane distance~\cite{gw}. 
In the following, this approach will be referred to as the stabilized RS approach.
One can further extend the previous setup by putting 
Standard Model gauge boson as well Standard Model 
fermions in the warped bulk (with the Higgs boson field being localized 
near the TeV brane) which is presently giving rise to a huge activity and,
in the following, will be referred to as the bulk RS approach.

Anticipating a bit on the following, extra dimensions model building becomes very
challenging especially in view of the present electroweak 
precision measurements and the experimental constraints from 
flavor physics.

The outline of this short overview naturally follows the above
discussion and will be divided into two main parts.
The first part will focus on the flat extra dimension models 
with the ADD, $TeV^{-1}$ and UED models.
The second part will be devoted to the warped extra dimension models
with minimal RS, stabilized RS and Bulk RS models.

Every possible models and signatures at the LHC will not be touched upon in this short 
overview because there are way too many for the size of this mini-eview but instead
some examples will be picked up to illustrate each topic mentioned in the outline. 

Higgsless models is discussed in~\cite{terning}. Black holes, string states, supersymmetric
extra dimensions models as well as models from intersecting branes (including intersecting 
branes at angle) will not be discussed in this short overview.

\section{Flat compactified extra dimensions models}

\subsection{ADD models}
As already mentioned, the ADD model is an approach where 
only gravity propagates in a bulk of $4 + n$ dimensions with 
$n$~compactified 
extra dimensions. The Standard Model fields are confined 
on a 4~dimensional brane. This model addresses the hierarchy 
problem since one can relate the ordinary 4~dimensional Planck scale $M^2_{Pl(4)}$ 
to a fundamental TeV scale via an extra-dimension volume factor $R^n$:
\begin{equation}
  M^2_{Pl(4)} = M^{n+2}_{Pl(4+n)} R^n 
\label{eq:add}
\end{equation}
where $R$ stands for the size of the $n$ 
compact extra dimensions.
For a fundamental scale $M_D=M_{Pl(4+n)}$ of the order of 1~TeV, very large extra 
dimensions can be expected. For example, in the case of one extra dimension, the 
size of the extra dimension can be of the order of the size of the solar system. 
Having in mind that the ADD approach also predicts important deviations from the 
Newton law of classical gravitation, one can see that the one extra dimension case 
is already ruled out experimentally since no subsequent effect of have been observed 
at the level of the solar system.
In the case of 2 or more extra dimensions the size can be of the order of the 
millimeter or nanometer. This scenario does not contradict submillimetric gravity 
measurements~\cite{adel} especially if the effects of the shape of the compactifying 
space is taken into account, even in the simplest case of toro\"{\i}dal 
compactifications~\cite{dienes}.

In addition to submillimetric gravity measurements one can also constrain the ADD 
scenario from various areas such as astrophysics and cosmology as well collider physics 
and explore its phenomenology. In the following the focus will be put on collider 
physics~\cite{addpheno} and more specifically the LHC. 
At colliders the production of KK~graviton states provides
the handle to sign the existence of compact extra dimensions.
One can have so called direct searches where KK~gravitons are 
present in the final state. In the ADD approach the KK~graviton states are close to each 
other in mass namely with mass differences down to a fraction of electron volt:
\begin{equation}
 \Delta m \sim ( { M_D \over TeV } )^{{n+2}\over 2} 10^{ {12n-31 \over n} }
\label{eq:grw}
\end{equation}
so that they form a quasi continuum of states which compensates
the smallness of their individual coupling to Standard Model field which
is of the order of the inverse of the usual 4~dimensional Planck mass $1/M_{Pl(4)}$.
At the LHC KK~gravitons can be produced in association with a jet from a quark
or a gluon, with a photon ($\gamma$) or a Z~boson thus giving rise to jet + missing energy, 
$\gamma$ + missing energy or Z + missing energy signature respectively (where the missing 
energy component is due to the escaping KK~graviton). The production cross sections are 
sizeable and directly related to the number of extra dimensions $n$ and to the scale $M_D$.
One can also have indirect searches where no KK~graviton states are present in the 
final states. Thus one has to look for deviations in fermion or boson pair production with 
respect to the prediction of the Standard Model.
In contrast to the direct searches, the cross sections are not directly 
related to the fundamental scale. For 2 and more extra dimensions the cross sections diverge and, 
in the context of field theory~\footnote{Cross section can be regularized in the context of type-I 
string theory.}, the introduction of a cut-off is required. However this cut-off is related to the
fundamental scale only through an arbitray factor which is usually taken equal to 1. 
Current collider constraints from HERA, LEP and Tevatron on scales are of the order of 
$1.6 - 2.1$~TeV for 2 extra dimensions.

\begin{figure}
\begin{center}
\begin{tabular}{cc}
\epsfig{figure=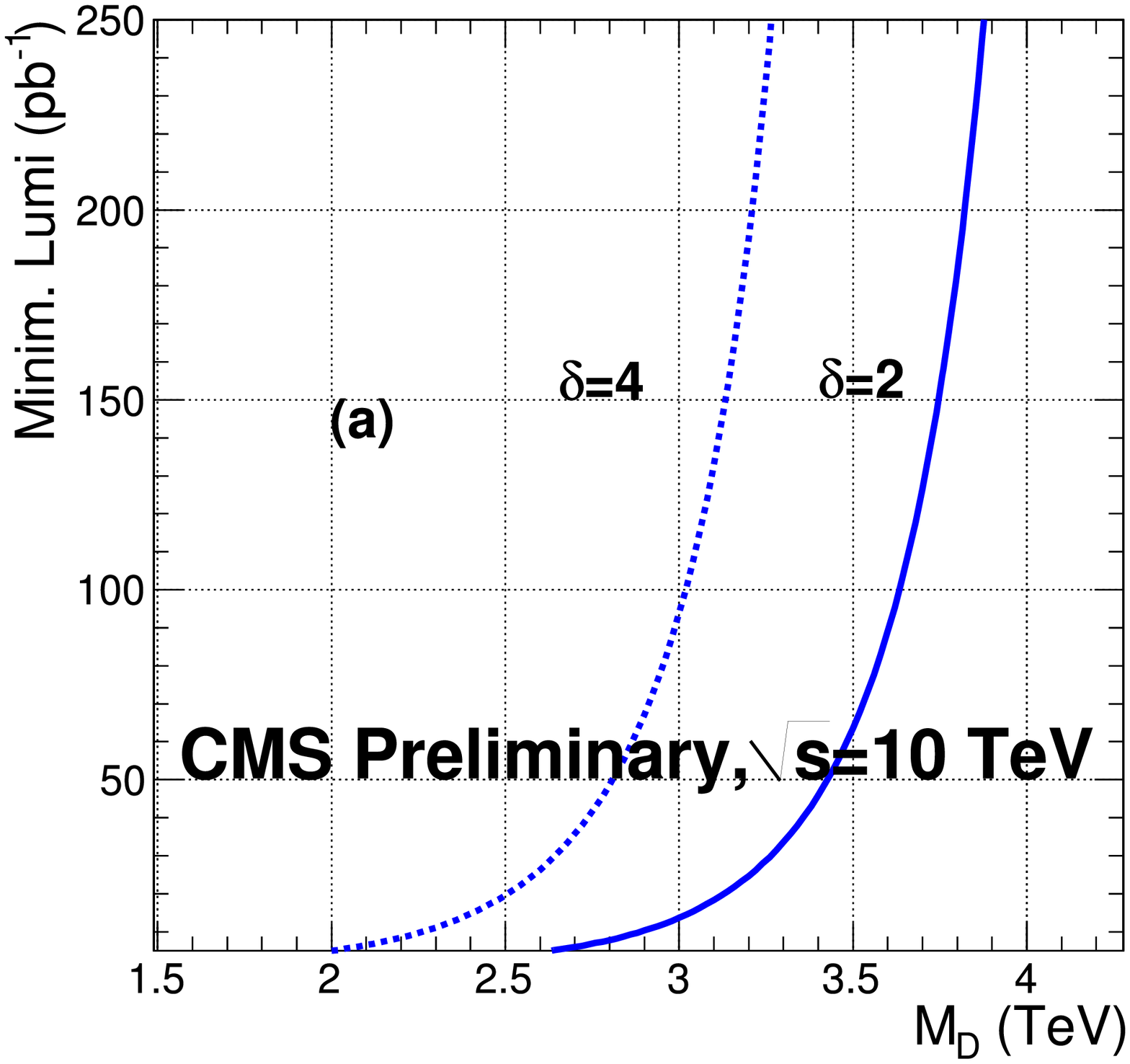,height=3.0in} &
\epsfig{figure=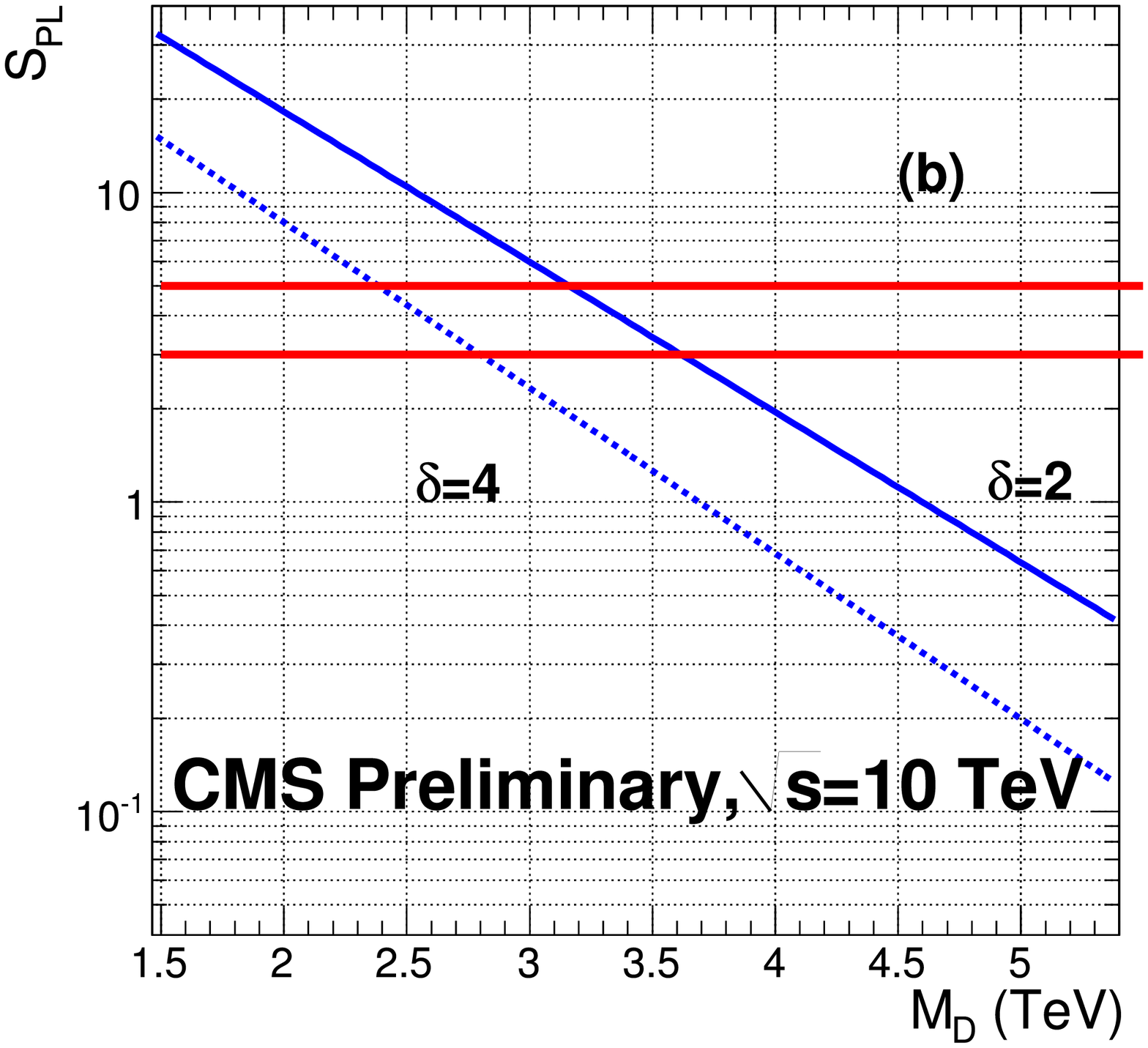,height=3.0in} \\
\end{tabular}
\end{center}
\caption{\it (a): exclusion at 95\% C.L. 
showing the minimum luminosity to exclude a given value of $M_D$. 
(b): discovery potential as a function of $M_D$ for various number
of extra dimensions $\delta$ after 200~pb$^{-1}$ of integrated luminosity.
The horizontal lines correspond to $3 \sigma$ and $5 \sigma$ significance
level. Both plots assume ${\sqrt s} = $10~TeV.
\label{fig:cmsadd}}
\end{figure}

At the LHC one can look at mono-jet events as done for example by the CMS
experiment~\cite{cmsadd1}. One can look for an excess of events over the background in 
observables such as the vectorial sum of jets $p_T$ after a simple set of cuts. Either exclusions 
above the current constraints or discovery can be achieved with relatively low luminosities i.e. 
O(100) pb$^{-1}$, as shown in Fig.~\ref{fig:cmsadd} from~\cite{cmsadd1}. Similar conclusions can 
be reached for indirect searches for example in the $\gamma \gamma$ channel as shown in~\cite{cmsadd2}.

\subsection{$TeV^{-1}$~Models}

As already mentioned TeV$^{-1}$ models are models where
the Standard Model gauge bosons can propagate in a
bulk with small compactified extra-dimensions of size
of the order of TeV$^{-1}$. For simplicity we can
here consider models with only one extra dimension.
In fact it has been shown that the 5~dimensional effective
gauge couplings are finite while for more than one extra dimension
they become divergent~\footnote{Again string theories and some brane 
configurations have to be invoked in order to regularize these couplings.}.
The standard model fermions are confined on a 4 dimensional brane and 
the KK 0$^{th}$ mode of the gauge bosons are identified with the 4 dimensional 
Standard Model gauge bosons.
A global fit including not only electroweak precision measurements but also
high energy data from LEP, HERA and the Tevatron Run~1 allows to set a lower 
limit of 6.8~TeV on the KK gauge bosons masses~\cite{tev-1}.
Direct searches at present colliders set lower limits of the order 
of 1~TeV. If kinematically allowed, KK gauge bosons can be resonantly produced 
at the LHC and one can look for a resonance decaying into fermions pairs.
Otherwise one has to rely on indirect effects and look for deviations in 
fermion pair production cross~sections measurements and asymmetries with
respect to the predictions of the Standard Model.
For example the study of the search for resonances
has been done in a quite generic way at the LHC
for both Z' and  W' types of gauge bosons. Depending on particular models 
and detector performances (lepton identification, missing energy resolution) 
one can achieve a $5 \sigma$ discovery from 1 TeV with O(10)~pb$^{-1}$ 
of well understood data up to 3~TeV where much more integrated luminosities are required 
namely in the O(10-100)~fb$^{-1}$ region as shown by the
ATLAS collaboration~\cite{atlastev-1}.

\subsection{Universal extra dimension (UED) models}
\begin{figure}
\begin{center}
\begin{tabular}{cc}
\epsfig{figure=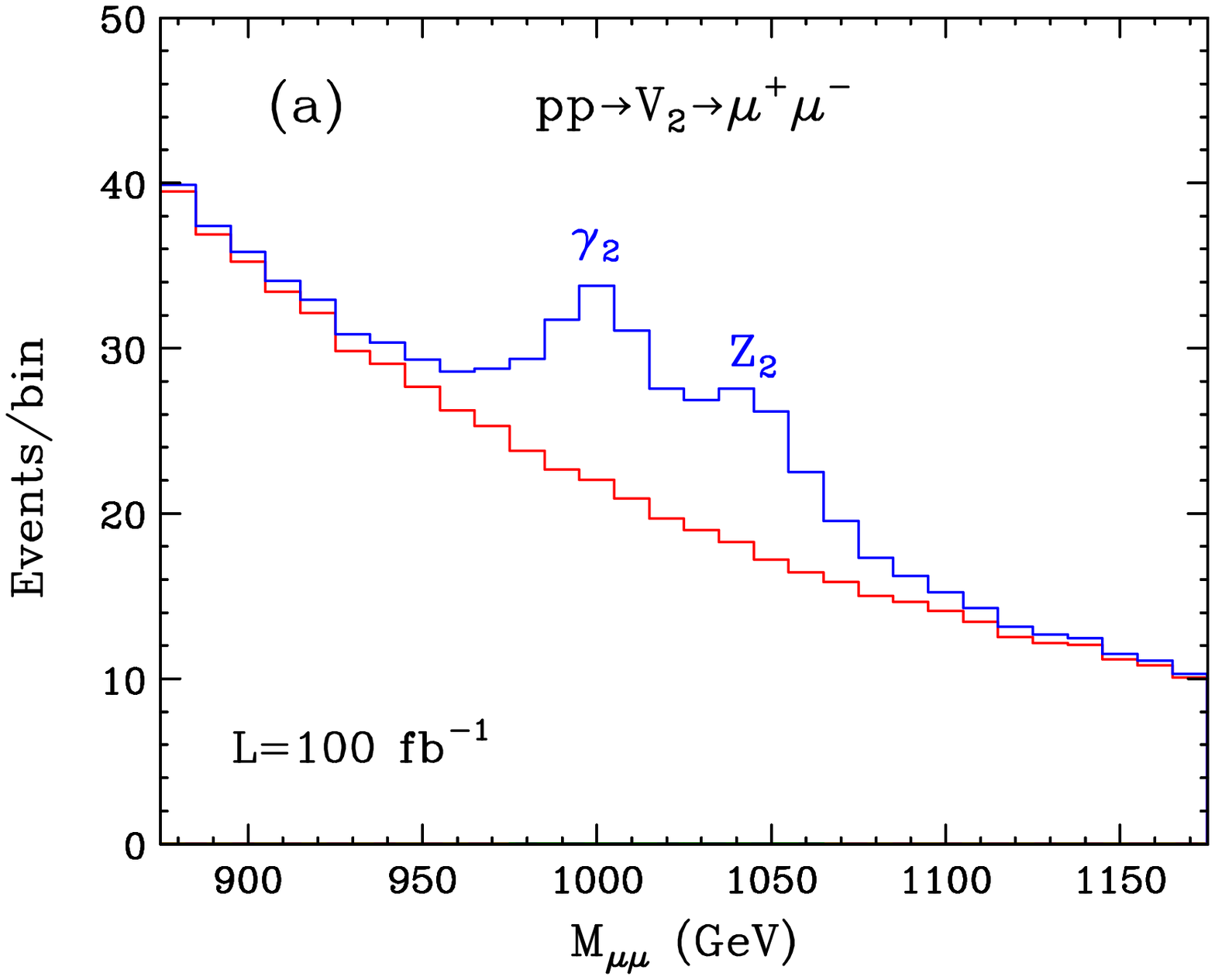,height=3.0in}
\epsfig{figure=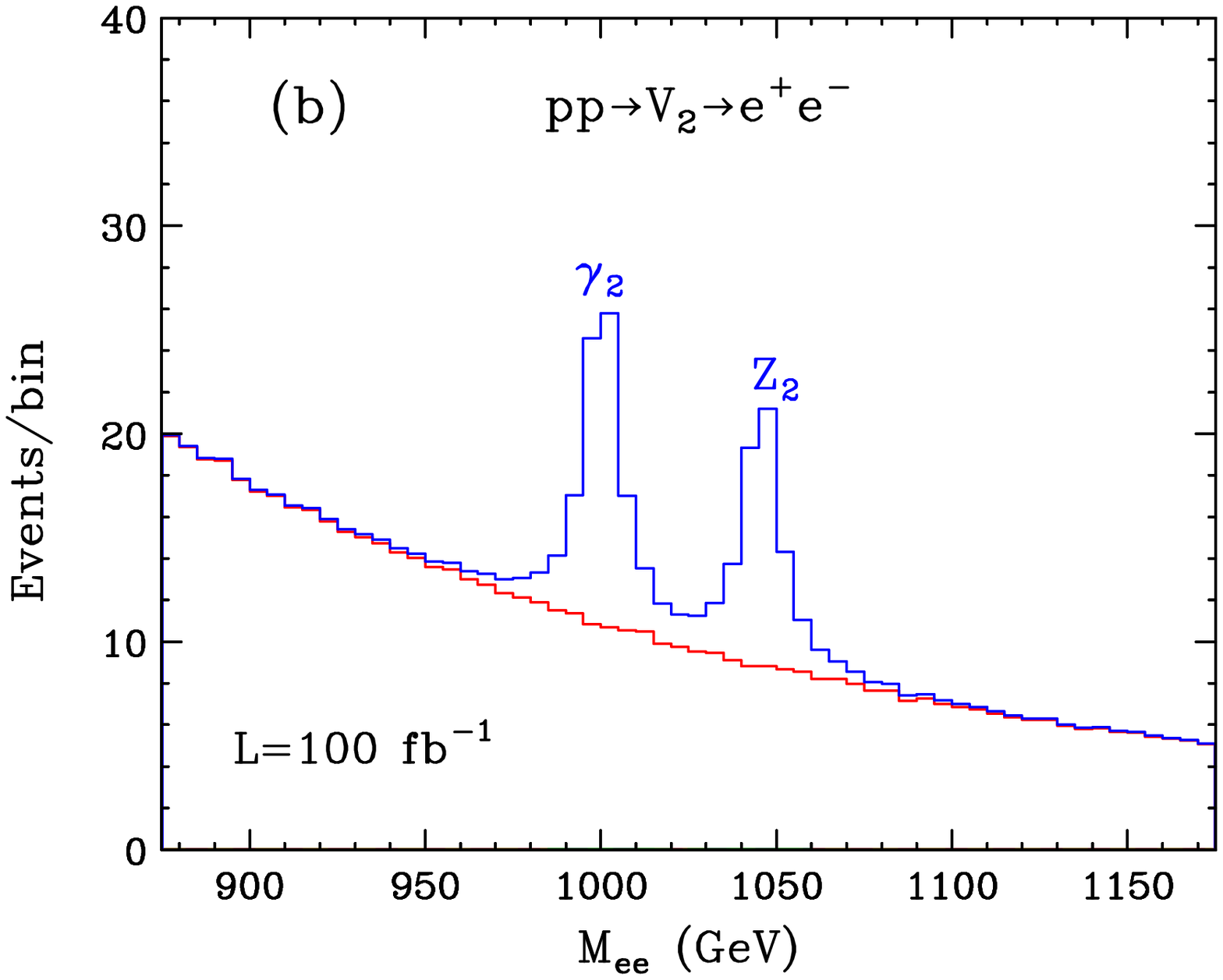,height=3.0in}
\end{tabular}
\end{center}
\caption{\it The $ \gamma_2 - Z_2 $ diresonance structure in UED (see text) with with $R^{-1}=500$~GeV
for the dimuons (a) and dielectron channel (b)  at the LHC with with 100~fb$^{-1}$ of integrated luminosity 
(at ${\sqrt s}= 14$~TeV)
.
\label{fig:ued}}
\end{figure}

UED models are models where 
all Standard Model fields propagate in the bulk (gravity is not included).
The minimal versions of such models are 
5 dimensional models with one extra compact 
dimension.
The KK 0$^{th}$ modes are identified with the 4 dimensional Standard Model 
particles. The non-zero KK modes are massive
and loop corrections involving bulk fields lead to non degenerate mass 
spectra~\cite{mat1,mat2}. The electroweak precision measurements set constraints 
on the typical mass scale $M$ of this UED scenario~\cite{ued} 
i.e. $M>300$~GeV. These constraints increase up to $M>700$~GeV when taking into 
account two-loop standard model contributions as well as LEP2 analyses~\cite{uednbound}.
Momentum conservation considerations in the bulk lead
to the conservation of a number call KK-parity
which in turn lead to a phenomenology resembling to the 
phenomenology of supersymmetry with conserved R-parity.
Namely  UED KK~states are pair produced, a UED~KK state 
decays into a UED KK~state and a particle of the Standard 
Model i.e. cascade decays can occur, and finally there exists 
a lightest KK~particle, the LKP, which is stable and which can 
escape a detector at colliders thus being a source of missing 
energy. The LKP provides a dark matter candidate which can be 
either the first KK mode of a photon $\gamma_1$ or the first 
KK~mode of a neutrino $\nu_1$~\cite{servant}.  
At the LHC the pair production of the lightest coloured KK~states 
have the largest production cross-sections~\cite{rizzoprobe}.
The production of UED KK~states, after cascade decays, can then 
lead to signatures such as multileptons and missing transverse 
energy or multi leptons , multijets and missing transverse energy
or multijets and missing transverse energy i.e. similar
to supersymmetry signatures.
As shown for example in~\cite{gigg}, in the 4 leptons and missing transverse 
energy channels,
one can expect a $5 \sigma$ discovery (at ${\sqrt s} = 14$~TeV) above
the current constraints with more than O(1)~fb$^{-1}$ of integrated
luminosity.
It would also be desirable to distinguish minimal UED signatures from 
supersymmetry signatures. In order to achieve this goal, one of the best 
way would be to search for the second level of KK~states, namely start to 
look for the KK tower structure. 
At similar masses the cross~sections of UED processes
are greater than the cross~section of supersymmetric processes (because both left and
right handed SU(2) doublet KK fermions are present in UED while one has only left
handed SU(2) squarks doublet, one integrates different angular distribution
i.e. $(1+\cos^2 \theta)$ for fermions versus ($1-\cos^2 \theta$) for scalars and
for production close to threshold one has different cross~section threshold
suppression i.e. $\beta$ for fermions versus $\beta^3$ for scalars). Level 2
KK~quarks can be directly pair produced (or produced in association with KK~gluons).
However cross~section times branching ratios for multi-leptons and missing transverse
energy channels, for example, are still challengingly small and there are challengingly small
statistics to distinguish from level~1 modes.
Alternatively the search for level~2 KK~gauge boson $V_2$ offers good prospect~\cite{kong} in 
particular when one includes the possibility of single KK~gauge boson production
via KK number violation interactions (but still with KK parity conservation)
i.e. looking for processes such as $pp \rightarrow V_2 \rightarrow f_0 {\bar f}_0$ where 
$f_0$ stands for fermion a 0$^{th}$ KK~level namely a fermion of the Standard Model.
One striking signature would be a double peak structure of $\gamma_2$ and $Z_2$ in 
dilepton invariant masses as shown in Fig.~\ref{fig:ued} with 
near mass degeneracy further corrobating the UED interpretation.
Preliminary studies~\cite{kong} show that one should be able to explore the 
O(0.6-1)~TeV mass range using from O(1)~fb$^{-1}$ up to O(100)~fb$^{-1}$
of integrated luminosity (at ${\sqrt s}= 14$~TeV).

\section{Warped extra dimension models}

\subsection{Minimal Randall Sundrum (RS) models}
As mentioned in the introduction, Randall and Sundrum (RS)~\cite{rs} proposed a phenomenological 
model with two 4~dimensional branes in a 5~dimensional space-time with a warped geometry
where gravity propagates (minimal RS). One can write the 
metric as $ds^2 = e^{-2k r_c \phi } \eta_{\mu\nu}dx^{\mu}dx^{\nu} + r_c d\phi^2$
with a compact 5$^{th}$ dimension
where $\phi$ sits in $[0,\pi]$ and with the warp factor $e^{-2k r_c \phi }$ where $k$
is a dimensionful parameter of the order of the 5~dimensional Planck scale $M_5$.
In contrast to the ADD relation (eq.~\ref{eq:add}), the 4~dimensional Planck scale
in the RS approach is given by:
\begin{equation}
 {M}^2_{Pl(4)} = { M^3_5 \over k } [ 1 - e^{-2kr_c\pi} ]  
\label{eq:rspl}
\end{equation}
so that $k$, $M_5$  and ${M}^2_{Pl(4)}$ have comparable magnitude when the warp factor is small.
The warp factor allows to generate 
a low energy scale on one brane (TeV brane), from a high energy 
scale, typically the Planck scale, on the other brane (the Planck brane). 
In particular a $\Lambda_{\pi} = M_{pl(4)} e^{-kr_c\pi}$=1~TeV energy scale can be generated 
from the 4~dimensional Planck scale if $k r_c \sim 12 $ ($r_c = 10^{-32}$~m)
thus allowing to solve the hierarchy problem. 
In contrast to the ADD approach the graviton field expansion
into KK modes in the RS approach is given by a linear combination of Bessel
functions. In consequence the masses of the KK~gravitons are not
regularly spaced but are given by $ m_n = x_n k e^{-k \pi r_c} $
where $x_n$ are the roots of Bessel functions. Furthermore
the order of magnitude of the mass of the first KK~graviton mode is O(1)~TeV. 
The Standard Model fields are localized on the TeV brane.
The coupling of the 0${^{th}}$ mode graviton to standard model fields
is inversely proportional to the 4~dimensional Planck mass and is thus suppressed. 
Nevertheless the coupling of the non~zero KK~graviton modes is inversely 
proportional to $e^{-k\pi r_c} M_{Pl}$ namely the 4~dimensional Planck mass 
multiplied by the warp factor.
i.e. the coupling of the non zero KK~graviton modes
to the standard model fields is enhanced by the warp factor $e^{k\pi r_c}$.
KK graviton modes are resonantly produced at colliders if they are kinematically accessible.
Once they are produced they decay
predominantly into two jets~\cite{hew} and then into other decay channels such as
$W^+W^-$, $ZZ$, $l^+l^-$, $t\bar t$ and $hh$.
Although leptonic decay channels are not dominant they offer a clear signature
in particular at the LHC. The phenomenomlogy of this minimal RS approach can be described
by two parameters namely the mass $M_1$ of the first KK~graviton mode and a parameter
$c=M_1/(x_1 \Lambda_{\pi})$ where typically $0.01 < c < 0.1$. Searches for the
KK~graviton have been performed at the Tevatron and constraints have been set~\cite{hays}.
By looking for a resonance decaying into
electron pairs it has also been shown that the
LHC should be able to cover the whole region
of interest with less than 100~fb$^{-1}$ at ${\sqrt s} = 14$~TeV
as shown in~\cite{atlasrs} and~\cite{cmsrs}.
It has also been shown that one should be able 
to distinguish between spin~1 and spin~2 resonances
by using angular distribution in the 
lepton lepton center of mass frame~\cite{alla}.

\subsection{Stabilized RS}
\begin{figure}
\begin{center}
\epsfig{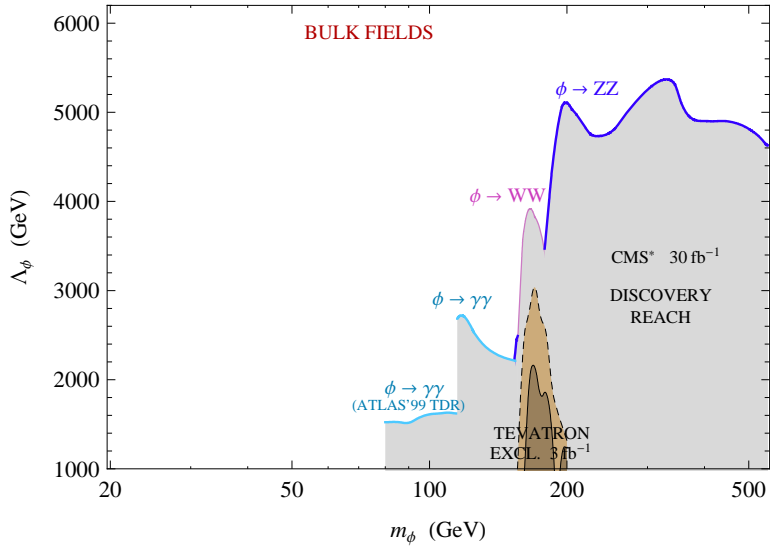}
\end{center}
\caption{\it Discovery reach for the radion at the LHC from Higgs boson 
searches from ATLAS and CMS (see text).
\label{fig:radion}}
\end{figure}

There are gravitational fluctuations around the
RS metric which contain a massless scalar mode called the radion.
The presence of this scalar field in the bulk with interactions
localized on the branes allows to stabilize the value of $r_c$
i.e. the interbrane distance~\cite{gw}. In order to recover 
ordinary 4 dimensional  Einstein gravity the radion must be massive
and, for this stabilized RS model to still solve the hierarchy problem,
the radion should be lighter than the KK~graviton. It turns out out that 
the radion is likely to be the lightest state from the RS models.
The radion couple to Standard Model fields via the trace of the 
energy-momentum tensor and can have direct coupling to gluon and photon.
The phenomenology of the radion resembles 
to the phenomenology of the Higgs boson for both 
production and decay except for this direct coupling
to gluon which allows to enhance the production with respect
to the Higgs boson and modify the light radion decay~\cite{radion}.
The radion predominantly decays into a gluon pair at low mass or W pair
above the WW mass threshold.
Besides, there also is, in addition, possible mixing between the Standard 
Model Higgs boson and the radion which allows to consider new physical mass 
eigenstates. The decay branching ratios of these new eigenstates are different 
from those of the Standard Model Higgs boson. Depending on the value of the 
coupling which is responsible of the Higgs boson-radion mixing the 
difference can be sizeable i.e. up to a factor 50 for the $W^+W^-$ et $ZZ$ 
decays. This mixing can also lead to non negligible
branching ratios for invisibly decaying Standard Model Higgs boson. This analysis has been 
confirmed in a more fundamental context involving type~I string theory~\cite{sturani}.
The Opal collaboration~\cite{opalrad} performed a search for the radion via existing searches
of the Higgs boson. No evidence for the radion has been found and the Opal collaboration derived
constraints on the parameters of the stabilized RS model (see~\cite{opalrad}) for various scenario
of Higgs-radion mixing.
Pure radion effects (i.e. without the above mentioned mixing) on precision 
electroweak observables have been shown to be small~\cite{toharia}.
It is possible to use the Standard Model Higgs boson searches
to search for the radion~\cite{usradion} as shown in Fig.~\ref{fig:radion} from 
M.~Toharia~\cite{tohariasusy}
where the explorable domain in the radion coupling and mass plane 
($\Lambda_{\phi} = {\sqrt 6} \Lambda_{\pi}$,$m_{\phi}$)
in a model where all Standard Model fields are in the RS bulk~\cite{hubisz} (anticipating a bit on the next
section).

\subsection{Bulk RS models}

Shortly after the development of the minimal and stabilized RS models, it has been realized 
that in order to solve the hierarchy problem of the Standard Model only the Standard Model
Higgs has to be localized near the TeV brane thus giving rise to the bulk RS models~\cite{bulkrs}.
In the bulk RS models, Standard Model fermions and gauge bosons are allowed to propagate 
in the RS bulk and the 4 dimensional Standard Model particles correspond to the 0$^{th}$ mode 
of the 5 dimensional fields. The bulk profile of the Standard Model fermion wave function depends
on its five dimensional mass parameter. The Yukawa coupling of Standard Model fermions depend
on the overlap of their wave function with the wave function of the Higgs boson. In other words
the masses and Yukawa couplings of the Standard Model fermions depend on the bulk profile of the
corresponding 5~dimensional fields. Thus bulk RS models allow to understand the Standard Model 
Yukawa coupling hierarchies and allow to suppress dangerous FCNC from higher dimensional couplings.
One can choose to localize the 1st and 2nd generation
fermions near the Planck brane where their wave functions have small overlaps between
with the wave function of the Higgs boson (localized near the TeV brane).
One can also choose also to localize the top and bottom quarks near the TeV brane with a bigger 
overlap of their wave functions with the wave function of the Higgs boson giving rise to bigger 
Yukawa coupling.
Electroweak precision data (including $Z \rightarrow b_L {\bar b}_L$) and data from flavor physics 
(K and B~physics, CP violation, rare decays) are very constraining for the bulk RS 
models~\cite{bulkrs}~\cite{contraintesbulkrs}. With the help of additionnal symmetries 
in the RS bulk such as custodial isospin symmetries sufficient to suppress
excessive contributions to the T parameter (and flavour symmetries) one can lower the lower
limits/constraints on KK masses namely O(3~TeV) for KK~gauge bosons, O(2-4~TeV) for KK~graviton
and  O(1-2~TeV) for fermions excitations. Without fermions in the bulk (i.e. having only gauge 
bosons in the bulk) and without custodial symmetries the lower limits on KK~graviton and KK~gauge
bosons would have been in the O(30-40~TeV) region i.e. well beyond the reach of the LHC.
Constraints from flavour physics are still striking hard any realistic model building attempt 
and a lot of effort is being put in RS flavor models developments~\cite{contraintesbulkrs} 
which are beyond the scope of this short overview.

There are many possible signatures for bulk RS models at the LHC~\cite{phenobulkrs}.
KK~gravitons (1st mode $G^{1}$) localized near the TeV brane can be resonantly and produced dominantly 
through the process $gg \rightarrow G^{(1)}$ for which the gluon profile in 
the bulk is supposed to be flat. The 1$^{st}$ generation fermions being localized near the Planck brane they have
small wave functions overlap with the KK~graviton and hence the KK graviton coupling to u~and d~quarks
is suppressed. At the LHC the production of KK~gravitons then occurs dominantly via gluons.
KK gravitons have decays which differ from the decays
of the minimal RS model i.e. KK gravitons decay predominantly into $t \bar t $, since the 3$^{rd}$ quark generation 
fermions is also supposed to be localized near the TeV brane, and into $WW$ and $ZZ$.
  
KK~gluons ($g^{(1)}$) and KK gauge bosons ($Z^{(1)}, W^{(1)}$) can also be resonantly produced with
$g^{(1)} \rightarrow t \bar t$, $Z^{(1)} \rightarrow WW $ and $W^{(1)} \rightarrow Wh$ decays.
Finally KK~fermions can be pair produced via for example $pp \rightarrow g+g^{(1)} \rightarrow t^{(1)} {\bar t}^{(1)}$.

Assuming for example a 10\% top identification efficiency, it has been shown that a 5~$\sigma$ reach (i.e. 
here in terms of $S/{\sqrt B}$ where S stands for the $gg \rightarrow G^{(1)} \rightarrow t \bar t$ signal and B for 
the background from  the Standard Model top quark pair production) can be achieved at the LHC 
(with ${\sqrt s} = 14$~TeV) for top quarks localized very near the TeV brane and for not too heavy KK gravitons 
i.e. in the O(1-2~TeV) mass range.

The top quarks from the KK~graviton decay are expected to be boosted. Improvements in the KK~graviton
mass reach are expected with the use of boosted top jet algorithms~\cite{taggedtop}
in order to improve the top tagging and identification efficiencies.
For example such algorithms have been used to search for KK~gluons~\cite{cmstoptagging} in 
all hadronic channels i.e. $pp \rightarrow g^{(1)} \rightarrow t {\bar t} \rightarrow b {\bar b} jjjj $. 
It has been shown that a typical 40 \% efficiency can be obtained for jet $p_T$ greater than 700~GeV
for a fake tag rate below 6 \% and that a $5 \sigma$ discovery can be made for cross~sections
of 43.6, 4, 1.6 and 1.3~pb respectively, this with 100~pb$^{-1}$ of integrated luminosity at 
${\sqrt s} =$~10 TeV.

\section{conclusion}

There is a wide spectrum of possible models and signatures of extra-dimensions
which can be explored at the LHC. There are strong constraints from electroweak 
precision data and data flavor physics which are very challenging for realistic
model building.
Some examples of non-resonant KK states searches with mono~jets 
as well searches for KK resonances in dileptons, di-boson and top quark pairs as well 
as the use of Higgs boson searches have been discussed.
Signal discovery can be achieved with relatively low integrated luminosities
considering optimistic model parameters and not too high KK~states masses.
However to establish that the discovered signals are actually coming from 
extra~dimensions and further dedicated studies may need a LHC running at the highest 
possible center~of~mass energy and  higher integrated luminosities.


\section*{Acknowledgments}
It is a pleasure to thank the organizers for the invitation to give this brief 
overview on models and signatures of 
extra dimensions at the LHC.

\end{document}